\pdfoutput=1
\documentclass[a4paper,11pt]{article}

\usepackage{contribution}

\newcommand{\weblink}[2][]{%
    \ifthenelse{\equal{#1}{}}%
    {\textnormal{\url{#2}}}%
    {\textnormal{\href{#2}{#1}}}%
}

\newcommand{\acknowledgements}[1]{%
  \bigskip\bigskip
  \textsf{\textbf{\Large Acknowledgements}} \\[2ex]
  {#1}
  \bigskip
}

\def\beq{\begin{equation}}
\def\eeq#1{\label{#1}\end{equation}}
\def\eeqn{\end{equation}}

\def\beqa{\begin{eqnarray}}
\def\eeqa#1{\label{#1}\end{eqnarray}}
\def\eeqan{\end{eqnarray}}

\let\bar=\overbar

\def\Dslash{\not{\hbox{\kern-4pt $D$}}}
\def\dslash{\not{\hbox{\kern-2pt $\del$}}}

\def\msb{{\bar{\ssstyle M \kern -1pt S}}}

\newcommand{\contribution}[7][]{%
  \clearpage
  \thispagestyle{plain}
  \ifthenelse{\equal{#1}{}}
  {\hypersetup{pdftitle={#2}}}
  {\hypersetup{pdftitle={#1}}}
  \hypersetup{pdfauthor={{#3} {#4}}}
  {\centering\normalfont\LARGE\bfseries\sffamily #2 \par\nobreak}
  \lhead{}
  \chead{%
    \textit{\footnotesize XIV International Conference on Hadron Spectroscopy
      (\weblink[\textit{hadron2011}]{http://www.hadron2011.de}), 13-17 June 2011, Munich, Germany}%
  }
  \rhead{}
  \bigskip
  \begin{center}
    {#3} {#4}\ifthenelse{\equal{#6}{}}{}{\footnote{\weblink[#6]{mailto:#6}}}
    \ifthenelse{\equal{#7}{}}{}{#7} \\
    \textit{#5}
  \end{center}
  \bigskip
}

\renewcommand{\abstract}[1]{%
  \begin{center}
    \begin{minipage}{0.85\textwidth}
      \begin{footnotesize}
        #1
      \end{footnotesize}
    \end{minipage}
  \end{center}
  \bigskip
}

\begin{document}

%
%
%
%
%
{  


\makeatletter
\@ifundefined{c@affiliation}%
{\newcounter{affiliation}}{}%
\makeatother
\newcommand{\affiliation}[2][]{\setcounter{affiliation}{#2}%
  \ensuremath{{^{\alph{affiliation}}}\text{#1}}}

\def\beq{ \begin{equation} }
\def\eeq{ \end{equation} }
\def\mystrut{\vrule height 2.0ex depth 2.0ex width 0pt}
\def\tallstrut{\vrule height 4.3ex depth 2.5ex width 0pt}
\def\tallstrutA{\vrule height 4.3ex depth 0ex width 0pt}
\def\deepstrut{\vrule height 4.2ex depth 2.6ex width 0pt}
\def\deepstrutA{\vrule height 0ex depth 2.6ex width 0pt}
\def\medstrut{\vrule height 1ex depth 1.0ex width 0pt}
\def\lowstrut{\vrule height 0ex depth 0.9ex width 0pt }
\def\lowstrutA{\vrule height 0ex depth 1.1ex width 0pt }
\def\upstrut{\vrule height 2.2ex depth 0.0ex width 0pt }
\def\Tcc{\hbox{$\,{T^{\pm}\kern-2.0ex\lowstrut}_{\bar c c}\,$}}
\def\Tbb{\hbox{$\,{T^{\pm}\kern-2.0ex\lowstrutA}_{\bar b b}\,$}}
\def\mythinspace{\kern0.05em}

%

\contribution[Heavy Baryons and Exotics Spectrum]  
{\strut\vskip-1.7cm Heavy Baryon Spectrum and New Heavy Exotics}  
{Marek}{Karliner}  
{\affiliation[Raymond and Beverly Sackler School of Physics and Astronomy,
 Tel Aviv University, Israel]{1} \\
 \affiliation[Department of Particle Physics,
Weizmann Institute of Science, Rehovot 76100, Israel]{2} \\
 \affiliation[High Energy Physics Division, Argonne National Laboratory
Argonne, IL 60439-4815, USA]{3} \\
\affiliation[Department of Physical Sciences,
University of Helsinki, POB 64, FIN-0014 Finland] {4}} 
{marek@proton.tau.ac.il}  
{\!\!$^,\affiliation{1}$, Harry J. Lipkin$\affiliation{2}\kern-0.10em^,\affiliation{3}$ and Nils~A.~T\"ornqvist\,\affiliation{4}}

\abstract{%
We discuss several highly accurate theoretical predictions for masses of baryons
containing the $b$ quark which have been recently confirmed by experimental data.
Several predictions are given for
additional properties of heavy baryons.
We also discuss the two charged exotic resonances $Z_b$ with quantum numbers of a
$(b \bar b u \bar d)$ tetraquark, very recently reported by Belle in the channel
$[\,\Upsilon(nS) \pi^+,\  n=1,2,3]$.
Among possible implications are deeply bound $I{=0}$ counterparts of the $Z_b$-s
and existence of a \,$\Sigma_b^+\Sigma_b^-$ dibaryon, a
\,{\em beauteron}.
}
%


\section{Introduction}
QCD describes hadrons as valence quarks in a sea of gluons and $\bar q q$ pairs.
At distances above $\sim 1~\hbox{GeV}^{-1}$ quarks acquire an effective
{\em constituent mass} due to chiral symmetry breaking.
A hadron can then be thought of as a bound state of constituent quarks.
In the zeroth-order approximation the hadron mass $M$ is then given by the sum of the masses of
its constituent quarks $m_i$,
\ $ M = \sum_i m_i\,.$
\ The binding and kinetic energies are ``swallowed" by the constituent quarks masses.
The first and most important correction comes from the color hyper-fine (HF)
chromo-magnetic interaction,
\begin{eqnarray}
\label{HFQCD}
\strut\qquad
M = \sum_i m_i + \sum_{i<j} V_{ij}^{HF(QCD)}\,;
\qquad
\qquad
V^{HF(QCD)}_{ij}  =
v_0 \,(\vec \lambda_i \cdot \vec \lambda_j)
\, {\vec \sigma_i \cdot \vec \sigma_j \over m_i m_j }
\, \langle \psi | \delta (r_i - r_j) | \psi \rangle
\end{eqnarray}
\noindent
where $v_0$ gives the overall strength of the HF
interaction, $\vec \lambda_{i,j}$ are the $SU(3)$ color
matrices, $\sigma_{i,j}$ are the quark spin operators and
$|\psi \rangle$ is the hadron wave function.
This is a contact spin-spin interaction, analogous to the EM hyperfine interaction,
which is a product of the magnetic moments,
\ $
V^{HF(QED)}_{ij} \propto \vec \mu_i \cdot \vec \mu_j =
e^2 \, \vec \sigma_i \cdot \vec \sigma_j /( m_i m_j )
$.\  In QCD, the $SU(3)_c$ generators take place of the electric charge.
From eq.~(\ref{HFQCD}) many very accurate results have been obtained for
the masses of the ground-state hadrons. Nevertheless, several caveats are
in order. First, this is a low-energy phenomenological model, still awaiting
a rigorous derivation from QCD. It is far from providing a complete description
of the hadronic spectrum, but it provides excellent predictions for mass
splittings and magnetic moments.
The crucial assumptions of the model are:
(a)
HF interaction is considered as a perturbation
which does not change the wave function;
(b)
effective masses of quarks are the same inside mesons and baryons;
(c)
there are no 3-body effects.
\section{Effective masses of quarks}
Constituent
quark mass differences depend strongly on the flavor of the spectator or
``neighbor"
quark \cite{KL2003}. For example,
$m_s-m_d \approx 180$ MeV when the spectator is a light quark but the same
mass difference
is only about 90 MeV when the spectator is a $b$ quark.
Since these are {\em effective masses},
we should not be surprised that their difference is affected by the
environment, but the large size
of the shift is quite surprising and its quantitative derivation from QCD
is an outstanding challenge for
theory.
We can extract the ratio of the constituent quark masses
from the ratio of the the hyperfine splittings in the corresponding mesons.
The hyperfine splitting between $K^*$ and $K$ mesons is given by
\beq
\label{KHF}
M(K^*){-} M(K) {=}
 v_0  { \vec \lambda_u \cdot \vec \lambda_s \over m_u m_s }
\left[ \left( \vec \sigma_u \cdot \vec \sigma_s\right)_{K^*}
{-}
\left( \vec \sigma_u \cdot \vec \sigma_s\right)_K \right]
\langle \psi | \delta(r) |\psi \rangle
{=}4 v_0 \, { \vec \lambda_u \cdot \vec \lambda_s \over m_u m_s }\,\langle \psi | \delta(r) |\psi \rangle,
\eeq
and similarly for hyperfine splitting between $D^*$ and  $D$
with \,$s \rightarrow c$ \,everywhere.
From (\ref{KHF}) and its $D$ analogue we then immediately obtain
\begin{eqnarray}
{M(K^*) - M(K)  \over M(D^*) - M(D) }
\approx {m_c \over m_s}
\end{eqnarray}
\subsection{Color hyperfine splitting in baryons}

As an example of hyperfine splitting in baryons,
let us now discuss the HF splitting in the $\Sigma\ (uds)$ baryons.
$\Sigma^*$ has spin ${3\over2}$, so the $u$ and $d$ quarks must be in a state
of relative spin 1. The $\Sigma$ has isospin 1, so the wave function of
$u$ and $d$  is symmetric in flavor. It is also symmetric in space,
since in the ground state the quarks are in a relative $S$-wave.
On the other hand, the $u$-$d$ wave function is antisymmetric in color,
since the two quarks must couple to a {\bf 3}$^*$ of color to neutralize the
color of the third quark. The $u$-$d$ wave function must be antisymmetric
in \hbox{flavor\,$\times$\,spin$\,\times$\,space\,$\times$\,color,} so it follows it must be
symmetric in spin, i.e. $u$ and $d$ are coupled to spin one. Since $u$ and $d$ are in spin 1 state
in both $\Sigma^*$ and $\Sigma$ their HF interaction with each other cancels between the two
and thus the $u$-$d$ pair does not contribute to the \,$\Sigma^*-\Sigma$\, HF splitting,
\beq
M(\Sigma^*) - M(\Sigma) =
6 v_0 \, { \vec \lambda_u \cdot \vec \lambda_s \over m_u m_s }\,\langle \psi | \delta(r_{rs}) |\psi \rangle
\label{HFSigma}
\eeq
\noindent
we can then use eqs. (\ref{KHF}) and (\ref{HFSigma}) to
compare the quark mass ratio obtained from mesons and baryons:
\beq
\left({{m_c}\over{m_s}}\right)_{Bar} =
{{M_{\Sigma^*} - M_\Sigma}\over{M_{\Sigma_c^*} - M_{\Sigma_c}}} = 2.84
;\qquad
\left({{m_c}\over{m_s}}\right)_{Mes} =
{{M_{K^*}-M_K}\over{M_{D^*}-M_D}}= 2.81
\eeq
\beq
 \left({{m_c}\over{m_u}}\right)_{Bar} =
{{M_\Delta - M_p}\over{M_{\Sigma_c^*} - M_{\Sigma_c}}} = 4.36
;\qquad
\left({{m_c}\over{m_u}}\right)_{Mes} =
{{M_\rho-M_\pi}\over{M_{D^*}-M_D}}= 4.46
\eeq
We find the same value from mesons and baryons $\pm2\%$.

The presence of a fourth flavor gives us the possibility of obtaining a new
type of mass relation between mesons and baryons. The $\Sigma - \Lambda$
mass
difference is believed to be due to the difference between the $u-d$ and
$u-s$
hyperfine interactions. Similarly, the $\Sigma_c - \Lambda_c$ mass
difference is believed to be due to the difference between the $u-d$ and
$u-c$
hyperfine interactions. We therefore obtain the relation
\beq
\left({
\displaystyle {1\over m_u^2} - {1\over m_u m_c}
\over
\displaystyle  {1\over m_u^2} - {1\over m_u m_s}}\right)_{\strut \kern-1ex
Bar/Mes}
\kern-4.0ex
=
\kern1em
{{M_{\Sigma_c} - M_{\Lambda_c}}\over{M_{\Sigma} - M_\Lambda}}=2.16
\phantom{aa}
\approx
\phantom{aa}
{{(M_\rho {-} M_\pi){-}(M_{D^*}{-}M_D)}
\over
{(M_\rho {-} M_\pi){-}(M_{K^*}{-}M_K)}}
{=}2.10\kern-3em\strut
\eeq
The meson and baryon relations agree to $\pm 3\%$.
We can
write down an analogous relation for hadrons containing the $b$
quark instead of the $s$ quark, obtaining the prediction
for splitting between $\Sigma_b$ and $\Lambda_b$:
\begin{equation}
{{M_{\Sigma_b} - M_{\Lambda_b}}\over{M_{\Sigma} - M_\Lambda}} =
{{(M_\rho - M_\pi)-(M_{B^*}-M_B)}\over{(M_\rho - M_\pi)-(M_{K^*}-M_K)}}= 2.51
\label{Sigma_b_pred}
\end{equation}
yielding
$M(\Sigma_b) - M(\Lambda_b) = 194 \,{\rm MeV}$ \cite{KL2003,Karliner:2006ny}.
This splitting was measured by CDF \cite{CDF_Sigma_b},
with isospin-averaged mass difference
$M(\Sigma_b) - M(\Lambda_b) = 192$ MeV.
There is also the prediction for the spin splittings, good to 5\%
\beq
M(\Sigma_b^*)- M(\Sigma_b) =
{{M(B^*)- M(B)}\over{M(K^*)-M(K)}}\cdot [M(\Sigma^*)-M(\Sigma)]=
22 \,{\rm MeV}
\eeq
to be compared with 21  MeV from the
isospin-average of CDF measurements \cite{CDF_Sigma_b}.
The challenge is to understand how and under what assumptions one can
derive from
QCD the very simple model of hadronic structure at low energies which leads
to such accurate
predictions.

\section{Magnetic Moments of Heavy Quark Baryons}

In $\Lambda$, $\Lambda_c$ and $\Lambda_b$ baryons the light quarks are
coupled to
spin zero. Therefore the magnetic moments of these baryons are determined
by the
magnetic moments of the $s$, $c$ and $b$ quarks, respectively. The latter
are
proportional to the chromomagnetic moments which determine the hyperfine
splitting
in baryon spectra. We can use this fact to
predict the $\Lambda_c$ and $\Lambda_b$ baryon magnetic moments by
relating them to the
hyperfine splittings in the same way as given in the original
prediction \cite{DGG} of the $\Lambda$ magnetic moment.
We obtain
\beq
\label{maglamc}
\mu_{\Lambda_c}= -2 {\mu_\Lambda}\cdot
{{M_{\Sigma_c^*} - M_{\Sigma_c}}\over{M_{\Sigma^*} - M_\Sigma}}
=0.43\phantom{7}\,{\rm n.m.};
\qquad
\mu_{\Lambda_b}= \phantom{-} {\mu_\Lambda}\cdot
{{M_{\Sigma_b^*} - M_{\Sigma_b}} \over{M_{\Sigma^*} - M_{\Sigma}}}
=-0.067 \,{\rm n.m.}
\end{equation}
We hope these observables can be measured in foreseeable future and
view the predictions (\ref{maglamc}) as a challenge
for the experimental community.

\section{Predicting the Masses of b-Baryons}
On top of the already discussed $\Sigma_b$ with quark content $bqq$, $q=u,d$.
there are two additional ground-state $b$-baryons, $\Xi_b$ and $\Omega_b$:

{\Large\boldmath$\Xi_b$\unboldmath:}
\ \ \ the quark content is $bsq$.
$\Xi_b$ can be obtained from an ``ordinary" $\Xi$ ($ssd$ or $ssu$)
by replacing one of the $s$
quarks by a $b$, with one important
difference.
In the ordinary $\Xi$, Fermi statistics
dictates that two $s$ quarks must couple to spin-1, while in the ground
state of $\Xi_b$ the $(sq)$ diquarks have spin zero.
Consequently, the $\Xi_b$ mass is given by the expression:
$
\Xi_b=m_b+m_s+m_u-3v\langle \delta(r_{us}) \rangle/{m_u m_s}
$.
The $\Xi_b$ mass can thus be predicted using the known $\Xi_c$ baryon mass
as a
starting point and adding the corrections due to mass differences and HF
interactions:
\begin{equation}
\Xi_b=\Xi_c + (m_b - m_c)
-{3v}
\left( \langle \delta(r_{us})
\rangle_{\Xi_b} -   \langle \delta(r_{us}) \rangle_{\Xi_c} \right)
/({m_um_s})
\eeq
Since the $\Xi_b$ and $\Xi_c$ baryons contain a strange quark,
and the effective constituent quark masses depend on the spectator quark,
the optimal way to estimate the mass difference $(m_b - m_c)$
is from mesons which contain both $s$ and
$b$ or $c$ quarks:
\beq
m_b - m_c =
\textstyle {1\over4}(3B_s^* + B_s) - {1\over4}(3D_s^* + D_s) =
3324.6 \pm 1.4~.
\label{eq_B_s_D_s}
\eeq
On this basis we predicted
\cite{Karliner:2007jp}
$M(\Xi_b) = 5795 \pm 5$~MeV. Our paper was submitted on June 14, 2007. The next day
CDF announced the result
\cite{Aaltonen:2007un},
$M(\Xi_b) =5792.9 \pm 2.5 \pm 1.7$ MeV,
following up on an earlier D0 measurement,
$M(\Xi_b) =5774 \pm 11 \pm 15$ MeV
\cite{Abazov:2007ub}.

{\Large\boldmath$\Omega_b$\unboldmath:}
\ \ \ for the spin-averaged $\Omega_b$ mass we have
\vskip-1.0cm
\beq
\label{Omegab-spin-ave}
\textstyle
{1\over3}
(2M(\Omega_b^*)+M(\Omega_b))=
{1\over3}
(2M(\Omega_c^*)+M(\Omega_c))
+{(m_b-m_c)\medstrut}_{B_s-D_s}
=6068.9\pm 2.4~\textnormal{MeV}
\eeq
\strut\vskip-1.0cm
\noindent
For the HF splitting we obtain
\strut\vskip-1.0cm
\beq
M(\Omega_b^*)-M(\Omega_b)=
(M(\Omega_c^*)-M(\Omega_c))\frac{m_c}{m_b}\frac{\langle \delta(r_{bs})
\rangle_{\Omega_b}}{\langle \delta(r_{cs}) \rangle_{\Omega_c}}=
30.7 \pm1.3~\textnormal{MeV}
\eeq
\strut\vskip-1.0cm
leading to the following predictions:
\strut\vskip-1.0cm
\begin{equation}
\label{omegab_pred}
M(\Omega_b) = 6052.1{\pm}5.6~\hbox{MeV};
\qquad\qquad
M(\Omega_b^*) = 6082.8{\pm}5.6~\rm{MeV}
\end{equation}
About four months after our prediction
(\ref{omegab_pred})
 for $\Omega_b$ mass
 \cite{otherb}, D0 collaboration published the first
measurement of $\Omega_b$ mass \cite{Abazov:2008qm}:
$M(\Omega_b)_{D0}=6165 \pm 10 (stat.) \pm 13 (syst.) \ \hbox{MeV}\,.$
The deviation from the central value of our prediction was huge,
113 MeV. Understandably, we were very eager to see the CDF result.
CDF published their result about nine months later,
in May 2009\cite{Aaltonen:2009ny}:
$M(\Omega_b)_{CDF}=6054 \pm 6.8(stat.) \pm 0.9(syst.) \ \hbox{MeV}\,.$
%
Fig.~\ref{b_baryons_TH_vs_EXP} shows a comparison of
our predictions for the masses of $\Sigma_b$, $\Xi_b$ and $\Omega_b$ baryons
with the CDF experimental data.
We have made additional predictions
\cite{Karliner:2007jp,otherb}
 for some excited states of $b$-baryons.
Our results are summarized in Table 10 of Ref.~\cite{otherb}.

\begin{figure}[t]
\centering
\includegraphics[width=0.70\columnwidth,angle=90]{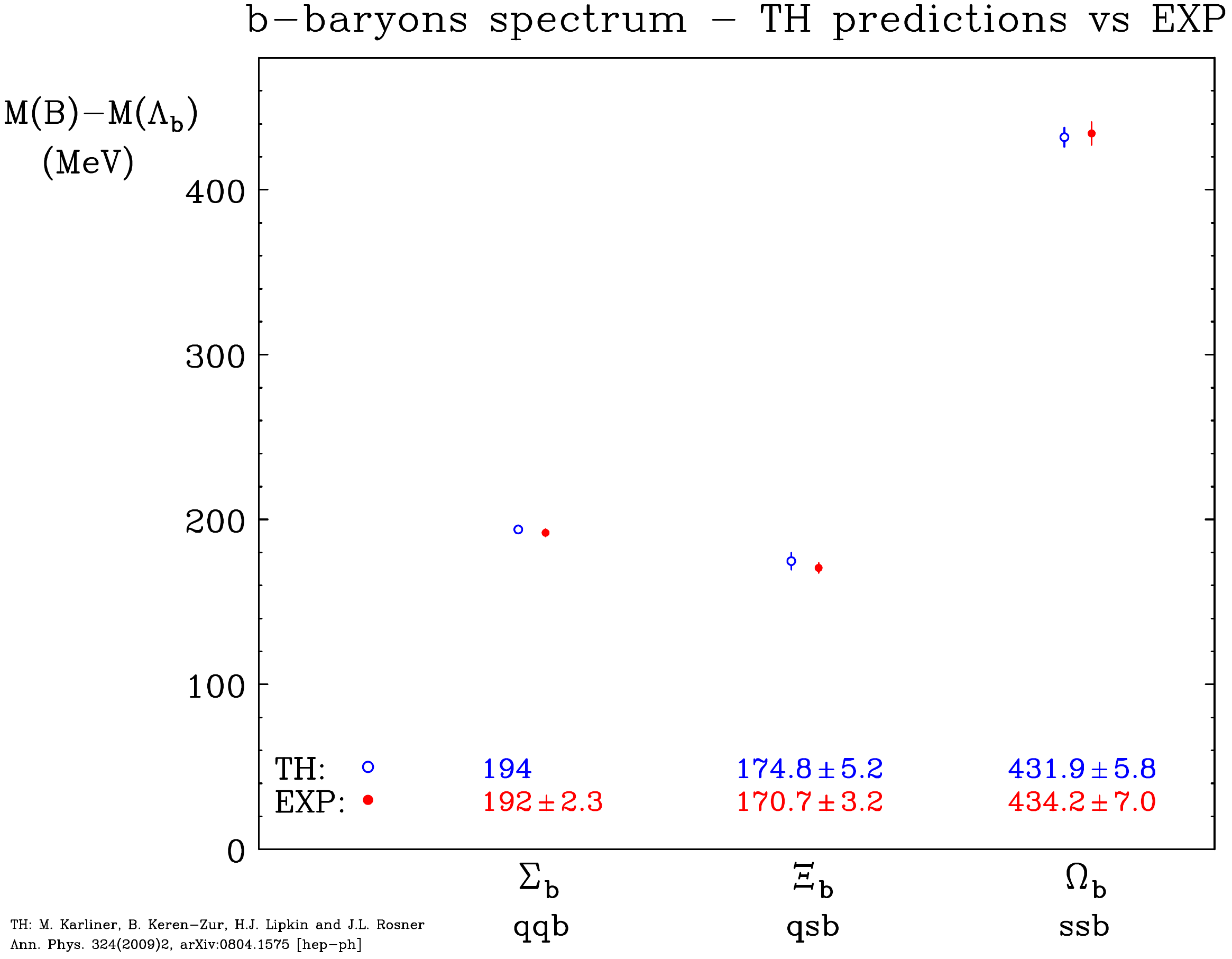}
\caption{Masses of $b$-baryons --
theoretical predictions
\cite{Karliner:2007jp,otherb}
vs. experiment.}
\label{b_baryons_TH_vs_EXP}
\end{figure}

The sign in our prediction
\ $M(\Sigma_b^*) - M(\Sigma_b) < M(\Omega_b^*) - M(\Omega_b)$,
 \ appears to be counterintuitive, since the color hyperfine
interaction  is inversely proportional to the quark mass.
This reversed inequality is  not predicted by other recent approaches
\cite{Ebert:2005xj,Roberts:2007ni,Jenkins:2007dm},
but it is also seen in the
charm data,
$M(\Sigma_c^*) - M(\Sigma_c)  =
64.3 \pm 0.5 \,\hbox{MeV} <  M(\Omega_c^*) - M(\Omega_c)
= 70.8\pm 1.5\, \hbox{MeV}$.
This suggests that the sign of the $SU(3)$ symmetry breaking gives information
about the form of the potential. It is of interest to follow this clue
theoretically and experimentally.

\section{Heavy exotics}
Ordinary hadrons contain either a $q \bar q$ pair or 3 quarks.
The possible color representations of quark combinations
are then completely determined by
confinement. In a meson the $q \bar q$~pair
{\em must} couple to a color singlet and
in a baryon any two quarks {\em must} couple to an anti-triplet of color,
to neutralize the color charge of the third quark.
The situation is very different in exotic hadrons
which contain both $qq$ and $q \bar q$ pairs,
eg. a tetraquark with two heavy quarks $Q$ and
two light quarks $q$,
\ $Q \bar Q  q \bar q$. Such states
have important color-space correlations that are completely
absent in ordinary mesons and baryons \cite{Karliner:2006hf}.
One also needs to keep in mind that the $q$-$\bar q$ interaction
is much stronger than $q$-$q$ interaction. The result is
emergence of
color structures that are totally different from those in
normal hadrons. In turn, this leads to some very unusual experimental
properties of such states. Until May 2011 the leading candidate
has been the $X(3872)$, which is most likely either a
$c \bar c \mythinspace q\bar q$
or a threshold bound state of $D$ and $\bar D^*$.
Given that $X(3872)$ exists, it is fascinating to explore possible
analogues containing $b$ quarks. General considerations suggest that such
states should be more strongly bound, since the attraction due to color
forces is the roughly same, but the repulsion due to kinetic energy
is smaller, as $E_k \sim p^2/m_Q$. Using a simple model, we have suggested
that
$b \bar b \mythinspace q\bar q$ might be below the $B \bar B$
threshold and
$b \bar c \mythinspace q\bar q$ might be below the $B \bar D$
threshold. A crucial difference vs. ordinary mesons is that
$(Qq)(\bar Q \bar q)$ can form a
$\bar{\hbox{{\bf 6}}}\mythinspace\hbox{{\bf 6}}$
color configuration which has much stronger binding than
$\bar{\hbox{{\bf 3}}}\mythinspace\hbox{{\bf 3}}$.
Some of these states have exotic electric charge, e.g.
$b d \mythinspace \bar c\kern0.07em \bar u \rightarrow J/\psi \pi^- \pi^-$.
Their decays have striking experimental signatures:
monoenergetic photons and/or pions, e.g.
$b q \mythinspace \bar c\kern0.07em \bar q$ with $\,I{=}0\,$ above $B_c \pi$
threshold can decay into $B_c \pi$ via isospin violation, or electromagnetically
into $B_c \gamma$, both very narrow.

\def\Jpsi{J\kern-0.12em/\kern-0.16em\psi}
Hadrons containing two $b$ quarks, such as double-bottom baryons $bbq$
or $b\bar b \mythinspace q \bar q$ and $b b \mythinspace \bar q \bar q$
tetraquarks have a unique and a spectacular decay mode with two $\Jpsi$-s
in the final state. To see this, recall that
a $b$ quark can decay via the hadronic mode
$b \rightarrow \bar c c s \rightarrow \Jpsi \mythinspace s$. If both $b$ quarks in
a double-bottom hadron decay this way,
for a $bb$ baryon we get
\ $(bbq) \rightarrow
\Jpsi \mythinspace \Jpsi (ssq) \rightarrow \Jpsi \mythinspace \Jpsi
\mythinspace\,\Xi$,
and similarly for a tetraquark:\quad
$(b\bar b \mythinspace q \bar q) \rightarrow \Jpsi \, \Jpsi (\bar s s \bar q q)
\rightarrow
\Jpsi \, \Jpsi \,K\, K,\ \hbox{etc.,}$
\ with all final state hadrons coming from the same vertex. This
unique signature is however hampered by a very low rate expected for such a process,
especially if one uses dimuons to identify the $\Jpsi$-s. It is both
challenge and a opportunity for LHCb \cite{Karliner:2006hf}.

\subsection*{Exotic double-bottom hadrons \boldmath $Z_b$: \unboldmath theoretical
prediction and discovery by Belle}
In 2008 Belle reported \cite{Abe:2007tk}
anomalously large (by two orders of magnitude)
branching ratios for the decays
\hbox{$\Upsilon(5S) \rightarrow \Upsilon(mS) \pi^+ \pi^-$,} \ $m=1,2$.
In \cite{Karliner:2008rc} we suggested that the enhancement is due to
an intermediate state of a tetraquark $T_{\bar bb} = (\bar b b u \bar d)$ and a pion,
mediating the two-step process
$$\Upsilon(5S) \, \rightarrow \, T_{\bar bb}^\pm
\,\pi^\mp \,\rightarrow\, \Upsilon(mS) \,\pi^+ \pi^-$$
We proposed looking for the $(\bar b b u \bar d)$
tetraquark in these decays as peaks in the
invariant mass of $\Upsilon(1S) \pi^+$ \,or \,$\Upsilon(2S) \pi^+$\, systems.

Very recently Belle collaboration confirmed this prediction, announcing
\cite{Collaboration:2011gja} the observation
of two charged bottomonium-like resonances $Z_b$ as narrow structures in
\hbox{$\pi^{\pm}\Upsilon(nS)$} \hbox{$(n=1,2,3)$} and \hbox{$\pi^{\pm}h_b(mP)$}
\hbox{$(m=1,2)$} mass spectra that are
produced in association with a single charged pion in $\Upsilon(5S)$ decays.

The measured masses of the two structures averaged over the five final
states are
\break
\hbox{$M_1=10608.4\pm2.0$ MeV,}
\,$M_2=10653.2\pm1.5$ MeV, both with a width of about 15 MeV.

Interestingly enough,
the two masses $M_1$ and $M_2$ are about 3 MeV above the
respective $B^* \bar B$ and $B^* \bar{B^*}$ thresholds.
This strongly suggests a parallel with $X(3872)$, whose mass is almost exactly
at the $D^* \bar D$ threshold.
It also raises the possibility that such states might have a complementary
description as deuteron-like ``molecule" of two heavy mesons quasi-bound by
pion exchange \cite{Tornqvist:1993ng,Thomas:2008ja}.

The attraction due to $\pi$ exchange is 3 times weaker
in the $I{=}1$ channel than in the $I{=}0$ channel.
This is because for $I{=}1$ only $\pi^0$ contributes,
whereas for $I{=}0$ both $\pi^0$ and $\pi^\pm$ contribute.
 Consequently, in the charm system the $I{=}1$ state is far above
the \,$D^* \bar D$\, threshold and only the $I{=}0$ \ $X(3872)$ is bound 2 MeV
below the average of the isospin-related $D^+D^{*-}$ and $D^0 \bar{D^{0}}$ thresholds.
The situation is likely to be different in the bottom system. This is because
the attraction due to $\pi$ exchange is essentially the same, but
the $B$ mesons are much heavier than $D$ mesons, so the kinetic energy is much smaller
by a factor of ${\sim} m(B)/m(D){\approx} 2.8$\,.
Therefore the net binding is much stronger than in the charm system. This raises
two very interesting possibilities:
\hfill\break
\def\mystrut{\vrule height 2.5ex depth 0ex width 0pt}
\mystrut
(a) the $Z_b$ states are almost bound (or quasi bound)
\,$B^* \bar B$\, and \,$B^* \bar{B^*}$\,\, $I{=}1$\,,$J^{P}=1^{+}$
states near threshold; the neutral members of their isomultiplets have $C{=}{-}1$,\,$G{=}{+}1$\,;
\,
\hfill\break
\mystrut
(b) since the binding in the \,$I{=}0$\, channel is much stronger than in the
\,$I{=}1$\,
channel, if we neglect effects other than $\pi$ exchange {\em we expect
the corresponding \,$I^G{=}0^+$\,,$J^{PC}=1^{{+}{+}}$\, states to be up to 40-50 MeV
\underline{below} the thresholds} \cite{KLT}.
The \,$I{=}0$\, states would then be expected close in mass to the $\Upsilon(4S)$.
Their expected decay modes are
$$ Z_b(I{=}0) \rightarrow \Upsilon(mS) \pi^+ \pi^-
\qquad\hbox{and}\qquad\, Z_b(I{=}0) \rightarrow \Upsilon(mS) \gamma\,,$$
as well as
$$ Z_b(I{=}0) \rightarrow B \bar B \gamma \quad \hbox{via} \quad
B^* \rightarrow B \gamma,\ \ E_\gamma = 46\  \hbox{MeV};$$
which might well be within the reach of LHCb.

\hfill\break
{\bf \boldmath A \,$(\Sigma_b^+\Sigma_b^-)$ \,{\em beauteron}\, dibaryon? \unboldmath}
\hfill\break
The discovery of the $Z_b$ states and their probable interpretation as
\,$B^* \bar B$\, and \,$B^* \bar{B^*}$\,
bound by pion exchange raises an interesting possibility that a
strongly bound $\Sigma_b^+\,\Sigma_b^-$ deuteron-like state might exist, a
{\em beauteron}.
This is because $\Sigma_b$ is about 500 MeV heavier than $B^*$ and
having $I{=}1$, it couples more strongly
to pions than $B$ and $B^*$ which have $I={1\over2}$. The opposite electric
charges of $\Sigma_b^+$ and $\Sigma_b^-$ provide an additional attraction.
A possible decay mode of the beauteron is
$$(\Sigma_b^+ \Sigma_b^-) \rightarrow \Lambda_b \,\Lambda_b \,\pi^+ \pi^-$$
which might be observable in LHCb. If the beauteron exists, it should also be seen
in lattice QCD.


\acknowledgements{%
The work on heavy baryons described
 here was done in collaboration with
B.~Keren-Zur and J.~Rosner.
It was supported in part by a grant from the
Israel Science Foundation.
The research of H.J.L. was supported in
part by the U.S. Department of Energy, Division of High Energy Physics,
Contract DE-AC02-06CH11357.}

\vfill\eject

%

}  



\begin{thebibliography}{99}

\bibitem{KL2003}
M. Karliner and H.J. Lipkin,hep-ph/0307243, Phys. Lett. {\bf B575} (2003) 249.

\bibitem{Karliner:2006ny}
  M.~Karliner and H.~J.~Lipkin,
  Phys.\ Lett.\  B {\bf 660}, 539 (2008)
  [arXiv:hep-ph/0611306].


\bibitem{CDF_Sigma_b}
T. Aaltonen {\em et al.} [CDF Collaboration], Phys. Rev. Lett. {\bf 99} (2007) 202001.


\bibitem{szmassqcd}
M.~Karliner and H.~J.~Lipkin,
  Phys.\ Lett.\  B {\bf 650}, 185 (2007)
  [arXiv:hep-ph/0608004].

\bibitem{DGG}{A. De Rujula, H. Georgi and S.L. Glashow, Phys. Rev. D12
(1975) 147}

\bibitem{KerenZur:2007vp}
  B.~Keren-Zur,
  Annals Phys.\  {\bf 323}, 631 (2008)
  [arXiv:hep-ph/0703011].

\bibitem{Karliner:2007jp}
  M.~Karliner, B.~Keren-Zur, H.~J.~Lipkin and J.~L.~Rosner,
  arXiv:0706.2163v1 [hep-ph].

\bibitem{Abazov:2007ub}
  V.~M.~Abazov {\it et al.} [D0 Collaboration],
  Phys.\ Rev.\ Lett.\ 99 (2007) 052001.

\bibitem{Aaltonen:2007un}
  T.~Aaltonen {\it et al.} [CDF Collaboration],
  Phys.\ Rev.\ Lett.\ 99 (2007) 052002.


\bibitem{otherb}
M.~Karliner, B.~Keren-Zur, H.~J.~Lipkin and J.~L.~Rosner,
  arXiv:0708.4027 [hep-ph] (unpublished) and
  arXiv:0804.1575 [hep-ph], Annals Phys {\bf 324},2 (2009).

\bibitem{Abazov:2008qm}
  V.~M.~Abazov {\it et al.}  [D0 Collaboration],
  Phys.\ Rev.\ Lett.\  {\bf 101}, 232002 (2008)
  [arXiv:0808.4142 [hep-ex]].

\bibitem{Aaltonen:2009ny}
  T.~Aaltonen {\it et al.}  [CDF Collaboration],
  Phys.\ Rev.\  D {\bf 80}, 072003 (2009)
  [arXiv:0905.3123 [hep-ex]].


\bibitem{Ebert:2005xj} D.~Ebert {\em et al.},
  Phys.\ Rev.\ D 72 (2005) 034026;
  Phys.\ Lett.\  B {\bf 659} (2008) 612.

\bibitem{Roberts:2007ni}
  W.~Roberts and M.~Pervin,
  arXiv:0711.2492 [nucl-th].

\bibitem{Jenkins:2007dm}
  E.~E.~Jenkins,
  Phys.\ Rev.\ D 77 (2008) 034012.

\bibitem{Karliner:2006hf}
  M.~Karliner and H.~J.~Lipkin,
  Phys.\ Lett.\  B {\bf 638}, 221 (2006)
  [arXiv:hep-ph/0601193].

\bibitem{Abe:2007tk}
  K.~F.~Chen {\it et al.}  [Belle Collaboration],
  Phys.\ Rev.\ Lett.\  {\bf 100}, 112001 (2008)
  [arXiv:0710.2577 [hep-ex]].

\bibitem{Karliner:2008rc}
  M.~Karliner and H.~J.~Lipkin,
  arXiv:0802.0649 [hep-ph].

\bibitem{Collaboration:2011gja}
  I.~Adachi {\it et al.} [Belle~Collaboration],
  arXiv:1105.4583 [hep-ex].


\bibitem{Tornqvist:1993ng}
  N.~A.~T\"ornqvist,
  Z.\ Phys.\  C {\bf 61}, 525 (1994)
  [arXiv:hep-ph/9310247];
%
  Phys.\ Lett.\  B {\bf 590}, 209 (2004)
  [arXiv:hep-ph/0402237].

\bibitem{Thomas:2008ja}
  C.~E.~Thomas, F.~E.~Close,
  Phys.\ Rev.\  {\bf D78}, 034007 (2008).
  [arXiv:0805.3653 [hep-ph]].

\bibitem{KLT}
M. Karliner, H.J. Lipkin and N.~A.~T\"ornqvist, unpublished.


\end{thebibliography}
\end{document}